# Photonic Dirac Nodal Line Semimetals Realized by Hyper-crystal

Shengyu Hu, Zhiwei Guo*, Haitao Jiang, and Hong Chen

*Key Laboratory of Advanced Micro-structure Materials, MOE, School of Physics Science and Engineering, Tongji University, Shanghai 200092, China*

**Abstract:** Recently, the gapless Dirac/Weyl nodal semimetals with linear dispersion and topologically protected modes degeneracy are rapidly growing frontiers of topological physics. Especially, type-I, type-II, and critical type-III nodal semimetals are discovered according to the tilt angles of the Dirac/Weyl cones. Here, by introducing hyperbolic metamaterials into one-dimensional photonic crystals, we design the "hyper-crystal" and study the photonic four-fold degenerate Dirac nodal line semimetals (DNLSs) with two types of perpendicularly polarized waves. Moreover, the flexibly controlled photonic DNLSs using the phase compensation effect of hyperbolic dispersion are studied. Our results not only demonstrate a new platform to realize the various photonic DNLSs, where the optical polarization plays the role of electron spin in electronic DNLSs, but also may pave a novel way to explore the abundant Dirac/Weyl physics in the simple classical wave systems.

* Corresponding author: 2014guozhiwei@tongji.edu.cn

## I. INTRODUCTION

According to the electronic band theory, materials can be classified mainly into metals, semiconductors and insulators. Since the 1980s, based on the incisive study of quantum Hall effect, people have put forward a new view, that is, topological order to describe the properties of matter [1-3]. By introducing the theory of topology, new bandgap phases, topological insulators and topological superconductors have attracted extensive attention [4], and led to special research on gapless topological phases such as topological semimetals in recent years [5-6]. In the momentum $k$-space, topological semimetals possess robust nodal areas between conduction and valence bands. In particular, the co-dimension of nodal points [7-11], nodal lines [12-41] and nodal surfaces [42-44] is 3, 2 and 1 respectively, which corresponding to the degrees of freedom of parameters that tuned to encounter a band degeneracy [45]. Interestingly, nodal lines in topological semimetals can even form some complicated configurations, including helices [12], rings [13-19], links [20, 21], chains [22-24], gyroscopes [18, 25-27], nexus [28-32], knots [33-36], nets [37-40], etc.

So far, more and more topological semimetals have been proposed theoretically and verified experimentally in the electronic [7-8], optical [11, 19, 24, 32], and acoustic systems [23, 39-41]. As a result, a finer classification of topological phases is supported by band structure of semimetals. A rapidly growing frontier in this field is to emulate relativistic quasi-particles like two-fold Weyl fermions [8, 41] and four-fold Dirac fermions [10, 19], which have linear dispersion with novel transport properties like delocalization [46], Zitterbewegung [11, 47-48], and Klein tunneling [11, 49-50]. In addition, topological semimetal phases can be further divided into type-I [51-52], type-II [53-54], and type-III [55-57] based on the tilt angles of the Dirac/Weyl cones. Especially, type-I and type-II have a point-like and conical-like Fermi surfaces, respectively. Type-III belongs to a critical transition phase, which is generally obtained by the topological phase transition (TPT) from type-I to type-II. Currently, several creative studies have realized this TPT in two-fold Weyl semimetals [55-61]. For the two-dimensional (2D) layered lattices, the tilt angle of the conical bands can be controlled by the compressive strain [55, 57, 59], lattice constant [56, 58, 61] and temperature [60]. All above mentioned modulations for the TPT resort to the change of coupling coefficient between neighbor unit cells, while the unit cells themselves remain unchanged, thus leading to the displacement of Weyl nodes in the process of TPT. This uncontrollable displacement will affect the observation of ideal TPT in photonic four-fold Dirac semimetals.

Photonic crystals (PCs) provide a powerful platform for manipulating the light-matter interactions [62-66]. In 2021, Hu *et al.*, uncover that double-bowl state in photonic Dirac nodal line semimetal in the one-dimensional (1D) PCs [19]. This pioneering work provides a new mechanism to realize a photonic type-II Dirac nodal line semimetal (DNLS). However, for the conventional 1D PCs composed of two types of dielectrics, the photonic bands will blueshift with the increase of the incident angle, thus the type-I, type-III DNLSs and even the TPT remain elusive in the 1D PCs. The recently emergent hyperbolic metamaterials (HMMs) have attracted people's great attention for their extraordinary optical properties, such as enhanced photonic density of states [73], abnormal coupling effect [74-76], unidirectional propagation [77-80], and so on. As one kind of anisotropic artificial media, HMMs can compensate the propagating phases to an unprecedented extent that usual dielectrics cannot attain [64], which provide the possibility to overcome the general angle-dependent limitation of bands and further explore the various DNLSs and TPT in 1D PCs [67-72].

This work is organized as follows: Sec. II covers the design of hyper-crystal composed of electric-magnetic (EM) HMM and dielectric, which satisfies the compensation condition $m_1 = m_2$. Especially, the photonic type-I DNLS is realized based on two type-I Weyl nodal line semimetals (WNLSs) of perpendicularly polarized waves. In addition, the TPT and type-III DNLS are demonstrated by tunning the thickness ratio of EM HMMs and dielectric layers; in Sec. III, hyper-crystal constructed by simple electric HMM and dielectric ($m_1 \neq m_2$) is carried out to study a hybrid photonic DNLS with a type-I WNLS and a type-II WNLS. Finally, Sec. IV summarizes the conclusions of this work.

## II. TPT and various photonic DNLSs realized by the hyper-crystal

DNLSs can be described by four-band Hamiltonian, which is expanded in Dirac gamma matrices. Considering the form of tensor product of two Pauli matrices $\tau_i$ and $\sigma_i$ ($i = x, y, z$) acting on two isospin degrees of freedom, the effective Hamiltonian of photonic DNLSs can be written as [19]

$$H_I = \tau_0 \otimes [(q_a - q_1)\sigma_z + (q_b - q_2)\sigma_x], \tag{1}$$

where $\tau_i$ and $\sigma_i$ represent the band index and pseudospin index, respectively. $\tau_0$ is the identity matrix. For the nodal line semimetals of co-dimension 2, $q_a$ and $q_b$ are the two compound variables of momentum, while $q_1$ and $q_2$ are constants. The effective Hamiltonian in Eq. (1) can be decoupled into two identical blocks, which provide two-fold Weyl nodal lines with different pseudospins α and β, respectively. The corresponding eigenvalues are given by

$$\begin{aligned} E_I^\alpha(\mathbf{k}) &= \pm\sqrt{(q_a - q_1)^2 + (q_b - q_2)^2}, \\ E_I^\beta(\mathbf{k}) &= \pm\sqrt{(q_a - q_1)^2 + (q_b - q_2)^2}. \end{aligned} \tag{2}$$

In order to understand the band structure, figure 1 shows the space of energy-momentum in a projection space $E - k_x - k_y$. We can see that the four-fold degenerate nodal line is a 2D manifold $S^2$: $\begin{cases} q_a = q_1 \\ q_b = q_2 \end{cases}$, which can be regarded as the path of Dirac cone running across the Brillouin zone. Moreover, constant $q_1$ and $q_2$ describe the size of shapes of closed manifolds or the location for open manifolds. Especially, there are some types of DNLSs that can be easily obtained. 1) $q_a = k_x$ together with $q_b = k_z$ support a straight nodal line. 2) $q_a = k_\rho = \sqrt{k_x^2 + k_y^2}$ together with $q_b = k_z$ support a nodal ring in the $k_x$-$k_y$ plane. 3) $q_a = k_\rho = \sqrt{k_x^2/C_1 + k_y^2/C_2}$ (constant $C_1, C_2 \in \mathbb{R}^+$) together with $q_b = k_z$ support a nodal elliptical ring in the $k_x$-$k_y$ plane. 4) $q_a = 2\cos k_x + 2\cos k_y$ together with $q_a = \sin k_x \cos k_z - \sin k_x \sin k_z$ support a double-helix nodal line [12]. Other configurations of DNLSs can be further obtained by deformation or combination with the above four situations, which means the potential correlation between various DNLSs.

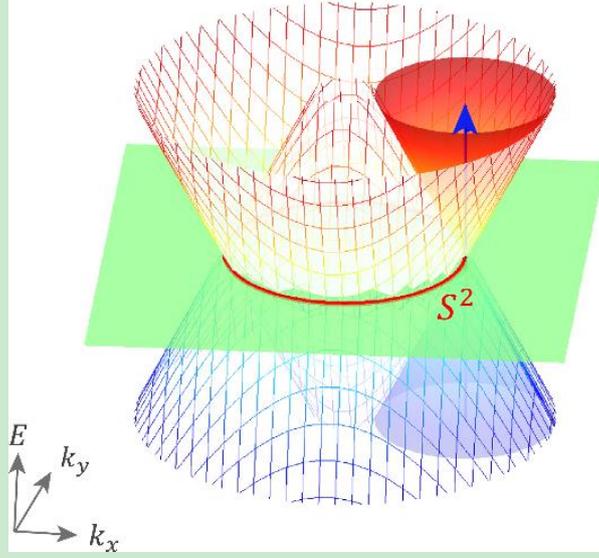

Fig. 1. The energy eigenvalues of DNLSs with effective Hamiltonian $H_I$. The zero-energy plane is represented by the green surface. In the space of $E - k_x - k_y$, the four-fold nodal line can be described by 2D manifold $S^2$ (red solid line). Nodal ring is realized based on the point-like Fermi surface (FS) running across the Brillouin zone.

We know that the effective Hamiltonian $H_I$ in Eq. (1) corresponds to a typical type-I DNLSs. After $H_I$ rotates around the Dirac point, the effective Hamiltonian becomes

$$H_{II} = \tau_0 \otimes \left\{ (q_a - q_1)[\tan(2\Phi)\,\sigma_0 + sec(2\Phi)\,\sigma_z] + \frac{(q_b - q_2)}{\sqrt{\cos(2\phi)}} \sigma_x \right\}, \qquad (3)$$

where $\phi$ denotes the rotation angle around the axis of $q_a$. The corresponding energy eigenvalues turn into

$$\begin{aligned} E_{II}^\alpha(\mathbf{k}) &= (q_a - q_1)\tan(2\phi) \pm \sqrt{\frac{(q_a-q_1)^2}{cos^2(2\phi)} + \frac{(q_b-q_2)^2}{\cos(2\phi)}}, \\ E_{II}^\beta(\mathbf{k}) &= (q_a - q_1)\tan(2\phi) \pm \sqrt{\frac{(q_a-q_1)^2}{cos^2(2\phi)} + \frac{(q_b-q_2)^2}{\cos(2\phi)}}. \end{aligned} \qquad (4)$$

According to Eq. (4), type-I and type-II DNLSs are realized for $\phi \in (n\pi - \pi/4, n\pi + \pi/4)$ and $\phi \in (n\pi + \pi/4, n\pi + 3\pi/4)$, which are shown in Fig. 2(a) and Fig. 2(d), respectively. In particularly, type-III DNLSs correspond to the critical rotation angle $\phi = n\pi \pm \pi/4$ ($n \in \mathbb{Z}$), as shown in Figs. 2(b) and 2(c). With the TPT from type-I to type-III and then to type-II DNLSs, the doubly degenerate Dirac cones gradually pass through the Fermi energy. In this process, the corresponding Fermi surface (FS) evolves from a point $(q_1, q_2)$ to a single line $q_b = q_2$ and then to two crossed lines $q_b -$

$q_2 = \pm\sqrt{-\cos(2\phi)}\,(q_a - q_1)$. This TPT of DNLSs has been proposed to support some exotic physical phenomena and applications, such as the black hole evaporation with high Hawking temperature [55].

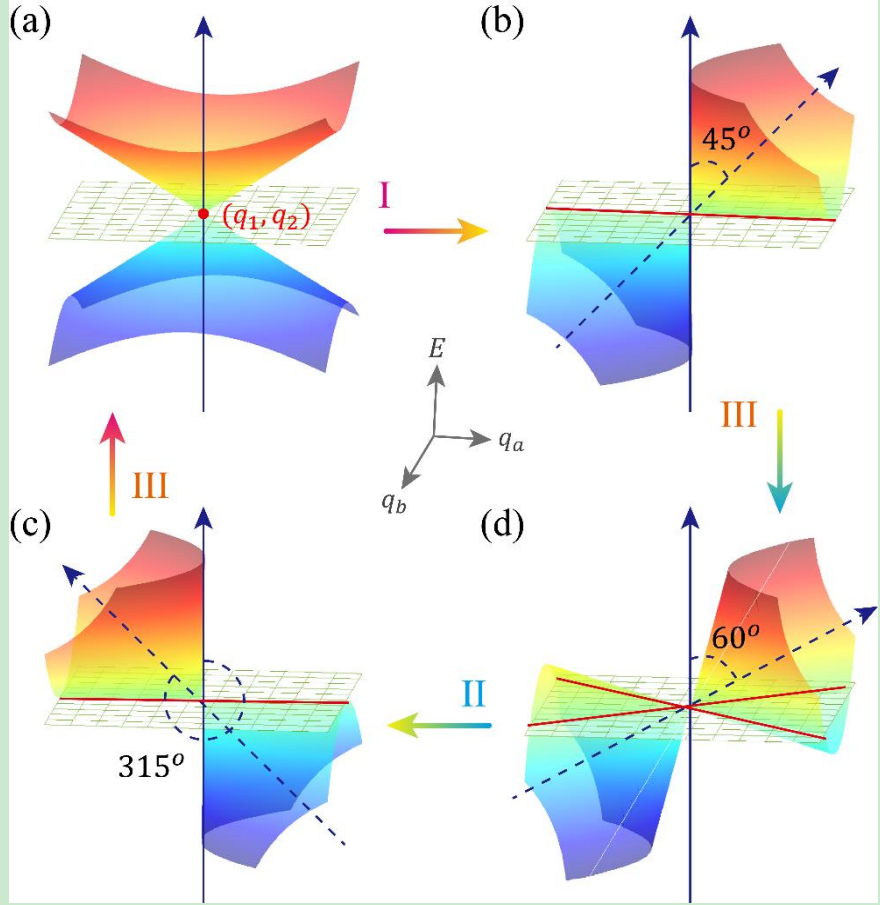

Fig. 2. Three types of DNLSs and the associated TPT. (a) The rotation angle of Dirac cone around axis $q_a$ is $\phi = 0^o$, which corresponds to type-I DNLSs with a point-like FS. (b)-(d) Similar to (a), but for the rotation angles are $\phi = 45^o$, $\phi = 60^o$, and $\phi = 315^o$, respectively. (b) and (c) correspond to type-III DNLSs with a line-like FS. (d) corresponds to type-II DNLSs with a conical-like FS. The arrows show the TPTs between three types of DNLSs. The zero-energy planes are represented by the mesh surfaces.

Then we study the above mentioned various DNLSs and the associated TPT in 1D PCs. We consider a 1DPC: $(AB)_N$ composed of an EM HMM layer $A$ and an isotropic material layer $B$, and the incident light propagates along the $z$ axis, as is shown in Fig. 3(a). The thickness of layers $A$ and $B$ is denoted by $d_A$ and $d_B$, respectively. Therefore, thickness of the unit cell is $\Lambda = d_A + d_B$. The permittivity and permeability tensors of two component layers are denoted by $diag(\varepsilon_{ix}, \varepsilon_{iy}, \varepsilon_{iz})$ and $diag(\mu_{ix}, \mu_{iy}, \mu_{iz})$ ($i = A\ or\ B$) respectively. The band degeneration of transverse-electric (TE, $H_x, E_y, H_z$) and transverse-magnetic (TM, $E_x, H_y, E_z$) polarized waves can be realized when the ratio of optical path between two component layers meets the conditions [19]:

$$\alpha^{TE} = \frac{\tilde{n}_A^{TE} d_A}{\tilde{n}_B^{TE} d_B} = \frac{m_1}{n_1} \in \mathbb{Q}, \text{ and } \alpha^{TM} = \frac{\tilde{n}_A^{TM} d_A}{\tilde{n}_B^{TM} d_B} = \frac{m_2}{n_1} \in \mathbb{Q}, \tag{5}$$

where $\{m_1, m_2, n_1\} \in \mathbb{N}^+$. In addition, the effective refractive index of two types of perpendicularly polarized waves can be expressed by $\tilde{n}_i^{TE} = \sqrt{\varepsilon_{iy}\mu_{ix} - \mu_{ix}k_\rho^2/\mu_{iz}k_0^2}$, and $\tilde{n}_i^{TM} = \sqrt{\varepsilon_{ix}\mu_{iy} - \varepsilon_{ix}k_\rho^2/\varepsilon_{iz}k_0^2}$, respectively. Especially, $k_\rho = \sqrt{k_x^2 + k_y^2} = k_0 \sin\theta$ is the wave vector along the radial direction, where $\theta$ is the incident angle, and $k_0 = \omega/c$ is the wave vector in a vacuum. $\omega$ and $c$ are the angular frequency and speed of light in a vacuum, respectively. For the TE waves, the $(m_1 + n_1)^{th}$ band and the $(m_1 + n_1 + 1)^{th}$ band intersect at the frequency $(f = \omega/2\pi)$

$$f_{m_1+n_1}^{TE} = f_0 = \frac{c(m_1+n_1)}{2(\tilde{n}_A^{TE} d_A + \tilde{n}_B^{TE} d_B)}. \tag{6}$$

Similarly, for the TM waves, the $(m_2 + n_1)^{th}$ band and the $(m_2 + n_1 + 1)^{th}$ band intersect at the frequency

$$f_{m_2+n_1}^{TM} = f_0 = \frac{c(m_2+n_1)}{2(\tilde{n}_A^{TM} d_A + \tilde{n}_B^{TM} d_B)}. \tag{7}$$

The ideal degeneration condition $m_1 = m_2$ corresponds to the photonic DNLS with $\phi_1 = \phi_2$ in the effective Hamiltonian $H_{II}$ in Eq. (3). According to Eq. (5), this ideal degeneration condition is equal to $\tilde{n}_A^{TE}(f_0) = \tilde{n}_A^{TM}(f_0)$ because $\tilde{n}_B^{TE} = \tilde{n}_B^{TM}$ is always satisfied for the isotropic material. Therefore, the EM HMMs will be considered as layer $A$ to realize the various photonic DNLSs. Especially, the EM HMM is mimicked by subwavelength $\varepsilon$-negative (ENG) media/ $\mu$-negative (MNG) media/dielectric stacks as $(CDE)_S$ in Fig. 3(a). $N = 50$ and $S = 8$ denote the period numbers of the PC and HMM, respectively. Both the permittivity of ENG media and the permeability of MNG media are described by Drude model [66]. The electromagnetic parameters and the thickness of different layers are shown in Table 1. Although this paper is a theoretical research work, two kinds of single-negative media have been widely studied by periodic arrays composed of meta-atoms [63, 81]. Therefore, the results studied in this work are feasible in experimental observation in the future.

|  | Permittivity ε | Permeability μ | Thickness (nm) |
|---|---|---|---|
| Layer B | 5.45 | 1 | 704 |
| Layer C | $3-5\times10^{30}/\omega^2$ | 1 | 16 |
| Layer D | 4 | $2-3\times10^{30}/\omega^2$ | 23 |
| Layer E | 9 | 1 | 117 |

**Table 1.** Electromagnetic parameters and thicknesses of different layers in the 1D hyper-crystal.

Based on the effective-medium theory (EMT) in the long-wavelength limitation, the effective electromagnetic parameters of layer $A$ as $(CDE)_S$ are given by [64]

$$\varepsilon_\| = \varepsilon_{Ax} = \varepsilon_{Ay} = \sum_j f_j \varepsilon_j, \quad \varepsilon_\perp = \varepsilon_{Az} = 1/\sum_j (f_j/\varepsilon_j),$$
$$\mu_\| = \mu_{Ax} = \mu_{Ay} = \sum_j f_j \mu_j, \quad \mu_\perp = \mu_{Az} = 1/\sum_j (f_j/\mu_j), \quad (8)$$

where $f_j = d_j/\sum_j d_j$ $(j = C, D, E)$ is the filling factor of the $j$ layer. Since the working wavelength is between 1363 nm and 1763 nm, which is much larger than the thickness of layer $A$: $d_A = d_C + d_D + d_E = 156$ nm, it is effective to use the EMT of Eq. (8). From the calculated effective parameters of layer $A$ in Fig. 3(b), it can be seen that the structure $(CDE)_S$ can be equivalent to an EM HMM ($\varepsilon_\|\varepsilon_\perp < 0$ and $\mu_\|\mu_\perp < 0$) when the frequency range is from 187 THz to 195 THz. $\varepsilon_\|$ ($\mu_\|$) and $\varepsilon_\perp$ ($\mu_\perp$) are represented by the red (blue) solid line and red (blue) dotted line, respectively. Combining Eq. (5) with Eq. (8), we can deduce that the ideal degeneration condition $m_1 = m_2$ of photonic DNLS is also equivalent to

$$\varepsilon_\|/\varepsilon_\perp = \mu_\|/\mu_\perp. \quad (9)$$

In order to intuitively determine the frequency of DNLS, $\varepsilon_\|/\varepsilon_\perp - \mu_\|/\mu_\perp$ is shown by the green dashed line in Fig. 3(b). Especially, the $\varepsilon_\|/\varepsilon_\perp = \mu_\|/\mu_\perp$ is marked by the green star at 190.4 THz. At this frequency, the effective refractive index of the TE (green solid line) mode $n_{Eeff}$ and TM (red triangles) mode $n_{Meff}$ are degenerate for different incident angles, as shown in Fig. 3(c). Therefore, the ideal degeneration condition of photonic DNLS in the 1D hyper-crystal is in good agreement with theoretical design at $f_0 = 190.4$ THz.

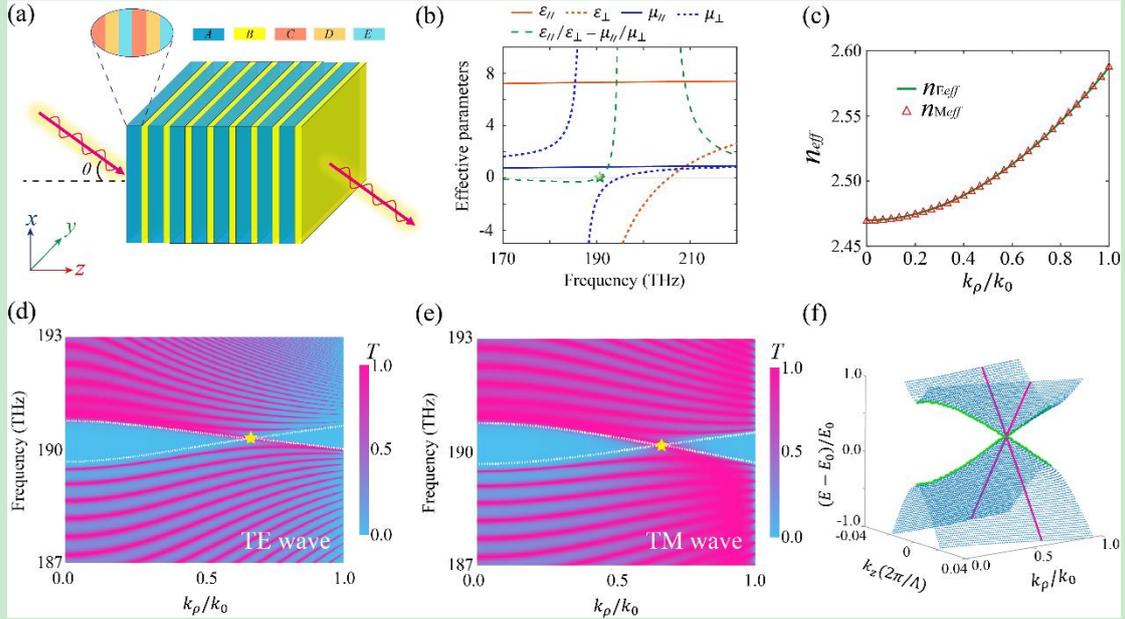

Fig. 3. (a) Scheme of a 1D hyper-crystal: $(AB)_N$ composed of an EM HMM layer A and an isotropic dielectric layer B. EM HMM is mimicked by subwavelength ENG/MNG/dielectric stacks as $(CDE)_S$. (b) Effective electromagnetic parameters of the structure $(CDE)_S$. $\varepsilon_{//}$ ($\mu_{//}$) and $\varepsilon_\perp$ ($\mu_\perp$) are represented by the red (blue) solid line and red (blue) dotted line, respectively. $\varepsilon_{//}/\varepsilon_\perp - \mu_{//}/\mu_\perp$ is denoted by the green dashed line and the ideal degeneration condition $m_1 = m_2$ is satisfied near the frequency of $f_0 = 190$ THz, which is marked by the green star. (c) At $f_0 = 190.4$ THz, the effective refractive index of structure $(CDE)_S$ as a function of $k_\rho/k_0$ for TE (green solid line) and TM (red triangles) waves. (d) The transmittance spectra of 1D hyper-crystal $(AB)_N$ with EMT: as a function of $k_\rho/k_0$ and the frequency for TE wave. The band edges with $k_z = 0$ are marked by white dotted lines. (e) Similar to (d), but for the TM wave. (f) Band structure of the 1D hyper-crystal in the normalized parameter space of $E - k_z - k_p$. The photonic type-I DNLS corresponds to $f_0 = 190.4$ THz and $k_\rho/k_0 = 0.66$. The linear dispersion is marked by the green and pink lines along directions of $k_\rho$ and $k_z$, respectively.

The phase variation compensation effect of HMM provides an effective manner to study the various photonic DNLSs [64]. Take the type-I DNLS for example, it can be obtained by

$$\frac{d_B}{d_A} = -\frac{\sqrt{\varepsilon_\| \varepsilon_B/\mu_\|}}{\varepsilon_\perp} \quad (|\varepsilon_\perp| \gg 1 \text{ and } \varepsilon_B \gg 1). \tag{10}$$

Using the transfer-matrix method, we calculate the transmittance spectra of the 1D hyper-crystal $(AB)_N$ with EMT for different normalized wave vector $k_p/k_z$. The results of the TE and TM modes are illustrated in Figs. 3(d) and 3(e) respectively. It can be clearly seen that the adjacent bands cross each other at $f_0 = 190.4$ THz and $k_\rho/k_0 = 0.66$ ($\theta = 40^o$), which is shown by the yellow star. In addition, the band edges with $k_z = 0$ are marked by white dotted lines. Band structure of the 1D hyper-crystal in the normalized parameter space of $E - k_z - k_p$ is shown in Fig. 3(f). The linear dispersions of the type-I DNLS along $k_p$ and $k_z$ directions are further demonstrated by the green and pink lines, respectively.

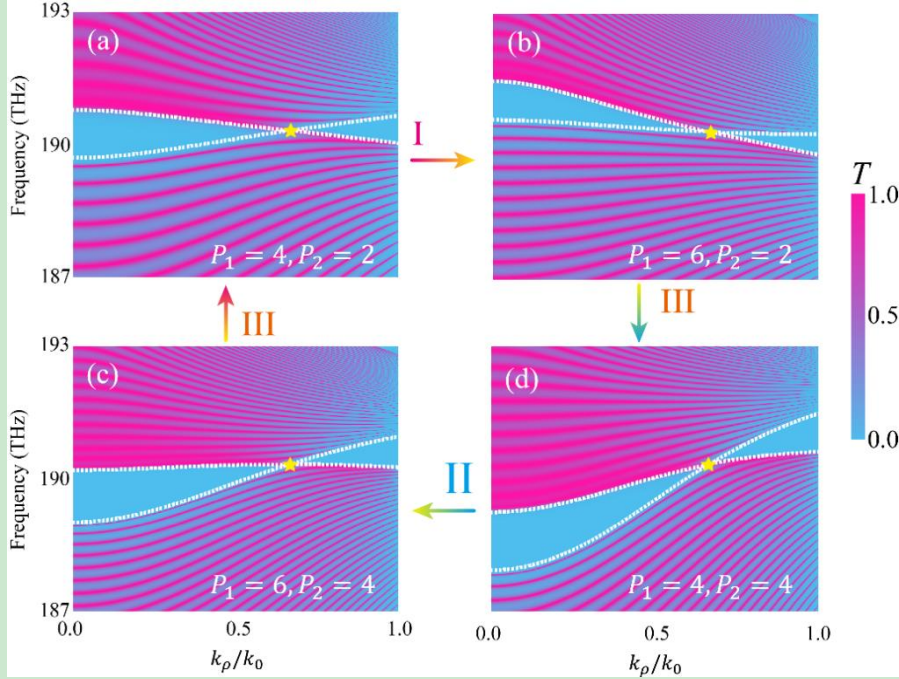

Fig. 4. For the TE waves, the transmittance spectra of 1D hyper-crystal $[(CDE)_S B]_N$ with different thickness factors: (a) $P_1 = 4$ and $P_2 = 2$ (Type I DNLS); (b) $P_1 = 6$ and $P_2 = 2$ (Type III DNLS); (c) $P_1 = 6$ and $P_2 = 4$ (Type III DNLS); (d) $P_1 = 4$ and $P_2 = 4$ (Type II DNLS). The various photonic DNLSs are marked by the yellow stars. The arrows show the TPTs between three types of DNLSs. The band edges correspond to $k_z = 0$, which are marked by the white dotted lines for see.

In the last part of this section, we introduce the TPT of photonic DNLSs in 1D hyper-crystal. From Eq. (5) we know that once the parameter group of thickness ($d_A$, $d_A$) can realize the DNLS at a degeneracy point (f, θ) with $m_1 = m_2$, all groups ($P_1 d_A/m_1$, $P_2 d_B/n_1$) will also degenerate at the same point. Here the thickness factor $P_1, P_2 \in \mathbb{Z}^+$ and the ratio of optical path in Eq. (5) becomes $P_1/P_2$. Especially, considering the effectiveness of Eq. (8), $P_1$ is modulated by changing the period number S of HMM layer A as $P_1 = S/2$. In addition, $P_2$ is controlled by changing the thickness $d_B$ of dielectric layer B as $P_2 = d_B/d_0$, where $d_0 = 352\ nm$. Combining the redshift property of HMM layer A and the blueshift property of dielectric B, the tilt angle of the degeneracy Dirac point of photonic DNLS can be flexible tuned and even reversed. For the 1D hyper-crystal with $m_1 = 2$ and $n_1 = 1$ (see more details in

Appendix C), the various photonic DNLSs can be realized in the process of TPT, as shown in Figs. 4 and 5. Figures 4 and 5 show the calculated transmittance spectra of 1D hyper-crystal with different thickness factors under TE and TM waves, respectively. First, the thickness factors are fixed at $P_1 = 4$ and $P_2 = 2$. The upright linear dispersions around the degenerate point ($f_0 = 190.4$ THz, $k_\rho/k_0 = 0.66$) for the TE and TM waves are shown in Figs. 4(a) and 5(a), respectively, which represent the type-I DNLS. The most obvious advantage of hyper-crystal over traditional PCs is that the type of DNLS, that is, the rotation angle of linear dispersion, can be flexibly changed by adjusting the ratio of $P_1$ to $P_2$. Next, when the thickness factors are fixed at $P_1 = 6$ and $P_2 = 2$, the linear dispersion rotates clockwise by an angle $\phi$ and becomes the type-III DNLS, as shown in Figs. 4(b) and 5(b). Furthermore, another type-III DNLS can also be realized at $P_1 = 6$ and $P_2 = 4$, which is shown in Figs. 4(c) and 5(c). Last, when $P_1 = 4$ and $P_2 = 4$, the rotation angle $\phi$ of linear dispersion for the TE and TM waves will further change, which corresponds to the type-II DNLS, as shown in Figs. 4(d) and 5(d), respectively. The TPTs between three types of DNLSs are shown by the arrows. It should be noted that, the various DNLSs realized in this section are based on the EM HMM with anisotropic electric and magnetic response. Based on the same study method, the optical WNLSs and the associated TPT can be easily realized in the 1D hyper-crystal with simple electric HMM or magnetic HMM [64, 70]. Nevertheless, this simple 1D hyper-crystal can also be used to study other novel physical properties. In the next section, we will systematically introduce the interesting hybrid photonic DNLS in the simple 1D hyper-crystal composed of electric HMM and dielectric.

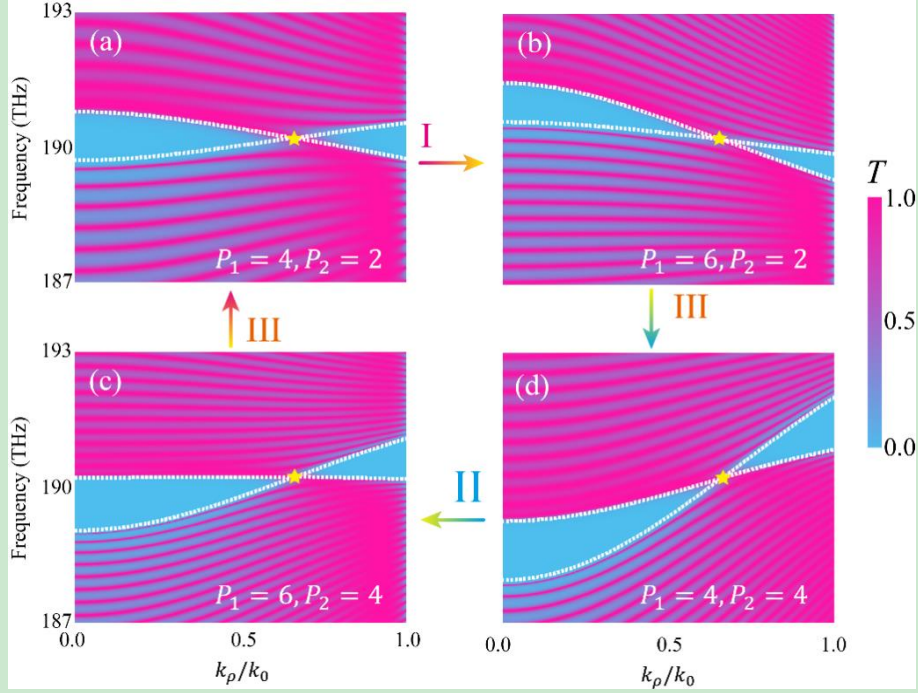

Fig. 5. Similar to Fig. 4, but for the transmittance spectra of the 1D hyper-crystal with different thickness factors under TM waves.

### III. Hybrid photonic DNLS realized by the hyper-crystal

In above section, the various photonic DNLSs have been demonstrated by the 1D hyper-crystal satisfying the ideal degeneration condition, which indicates the identical rotation angle for TE and TM waves. In fact, the rotation angle of linear dispersion for different pseudospins can also be tuned differently as $\phi_1 \neq \phi_2$ In this case, the effective Hermitian model in Eq. (3) can be expressed as

$$H_{III} = \tau_0 \otimes \{X + Y\} + \tau_z \otimes \{M + N\}, \tag{11}$$

where $X = (q_a - q_1)(a^+\sigma_0 + b^+\sigma_z)$, $Y = (q_b - q_2)c^+\sigma_x$, $M = (q_a - q_1)(a^-\sigma_0 + b^-\sigma_z)$, and $N = (q_b - q_2)c^-\sigma_x$. In addition, $a^\pm = [\tan(2\phi_1) \pm \tan(2\phi_2)]/2$, $b^\pm = [\sec(2\phi_1) \pm \sec(2\phi_2)]/2$, and $c^\pm = \frac{1}{2\sqrt{\cos(2\phi_1)}} \pm \frac{1}{2\sqrt{\cos(2\phi_2)}}$. The eigenvalues of Eq. (11) are given by

$$\begin{aligned} E_{III}^\alpha(\mathbf{k}) &= (q_a - q_1)\tan(2\phi_1) \pm \sqrt{\frac{(q_a-q_1)^2}{\cos^2(2\phi_1)} + \frac{(q_b-q_2)^2}{\cos(2\phi_1)}}, \\ E_{III}^\beta(\mathbf{k}) &= (q_a - q_1)\tan(2\phi_2) \pm \sqrt{\frac{(q_a-q_1)^2}{\cos^2(2\phi_2)} + \frac{(q_b-q_2)^2}{\cos(2\phi_2)}}. \end{aligned} \tag{12}$$

According to Eq. (12), the hybrid DNLS composed of two WNLSs with different semimetal phases can be obtained, as shown in Fig. 6(a).

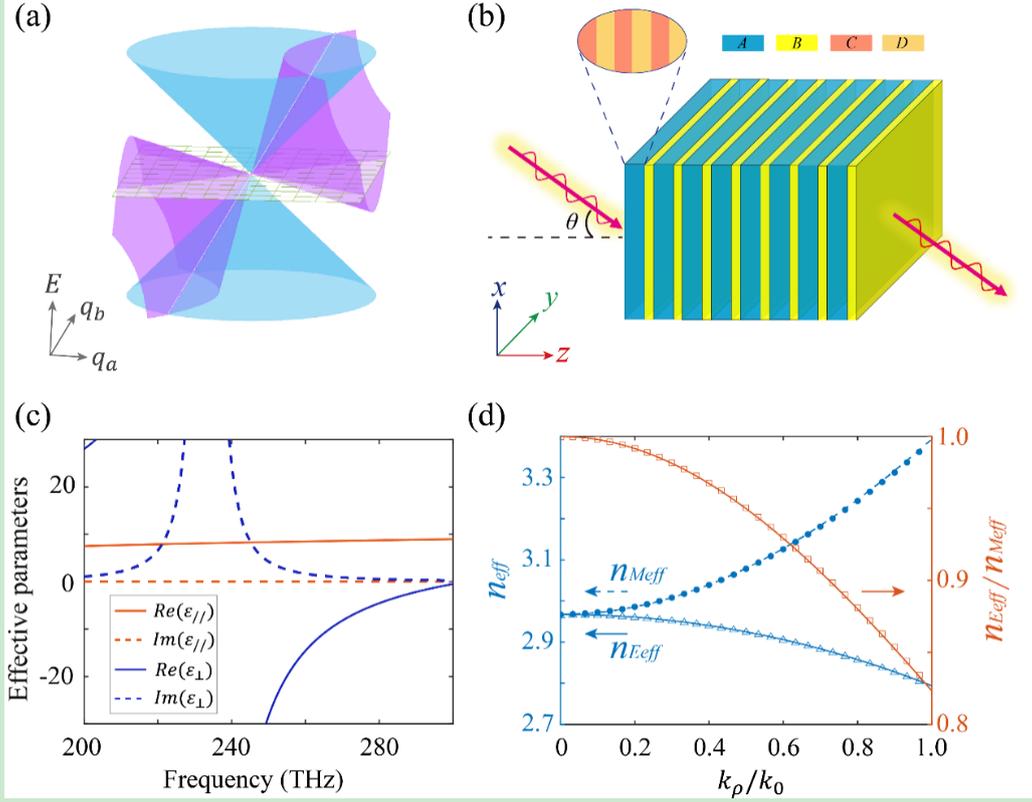

Fig. 6. (a) The energy eigenvalues of hybrid DNLS with effective Hamiltonian $H_{III}$. In the space of $E - q_a - q_b$, two Dirac cones for different pseudospins degenerate at one point, which means the hybrid DNLS. (b) Scheme of a 1D hyper-crystal: $(AB)_N$ composed of an electric HMM layer $A$ and an isotropic dielectric layer $B$. Electric HMM is mimicked by subwavelength metal/dielectric stacks as $(CD)_S$. (c) The corresponding effective electromagnetic parameters of the structure $(CD)_S$. Real part and imaginary part of $\varepsilon_{//}$ ($\varepsilon_\perp$) are represented by the red (blue) solid line and dashed line, respectively. (d) The effective refractive index of the TE wave: $n_{Eeff}$ (blue line consisting of circle dots), TM wave: $n_{Meff}$ (blue line consisting of triangles) and the ratio of $n_{Eeff}$ and $n_{Meff}$ (orange line consisting of squares) on $k_\rho/k_0$ at the frequency of $f_0 = 285$ THz.

The 1D hyper-crystal $(AB)_N$ composed of electric HMM layer $A$ and dielectric layer $B$ can be used to realize the hybrid photonic DNLS. The corresponding scheme is shown in Fig. 6(b). Here the electric HMM is mimicked by the widely used subwavelength metal/dielectric stacks as $(CD)_S$ [67-70]. The indium tin oxide (ITO) is selected as the metal layer $D$ and the corresponding permittivity is described by the Drude model [82]

$$\varepsilon_D(\omega) = \varepsilon_\infty - \frac{\omega_{pD}^2}{\omega^2 + j\omega\gamma_D}, \quad (13)$$

where $\varepsilon_\infty = 3.9$ is the high-frequency permittivity. In addition, $\hbar\omega_{pD} = 2.48$ $eV$ and $\hbar\gamma_D = 0.016$ $eV$. $\omega_{pC}$ and $\gamma_D$ denote the plasma frequency and damping frequency, respectively. Potassium titanyl phosphate (KTiOPO₄, KTP) [83] and Zinc silicon arsenide (ZnSiAs$_2$) [84, 85] are selected for dielectric layers *B* and *C*, respectively. The refractive indexes of layer *B* and layer *C* are close to 1.47 and 3.3, and there is slight dispersion at the working frequency (see Appendix C for details). Because all of the materials are non-magnetic, the permeability of each layer in Fig. 6(b) is $\mu_B = \mu_C = \mu_D = 1$. In addition, $N = 10$ and $S = 15$ denote the period numbers of the PC and electric HMM, respectively. The thicknesses of *B*, *C* and *D* layers are $d_B = 2352$ nm, $d_C = 68$ nm, and $d_D = 17$ nm, respectively. According to EMT, the effective anisotropic permittivity of the structure $(CD)_S$ is shown in Fig. 6(c). Real part and imaginary part of $\varepsilon_{//}$ ($\varepsilon_\perp$) are represented by the red (blue) solid line and dashed line, respectively. When the frequency ranges from 233 THz to 300 THz, $\varepsilon_{//}\varepsilon_\perp < 0$, which indicates that layer *A* is an effective electrical HMM [67-70]. Similar to Fig. 3(c), the effective refractive indexes of layer *A* for TE and TM waves on $k_\rho/k_0$ are shown by blue circle dots line and blue triangles line in Fig. 6(d). In order to see the polarization dependent properties of the 1D hyper-crystal, the ratio of $n_{Eeff}$ and $n_{Meff}$ is also given by the orange squares line in Fig. 6(d).

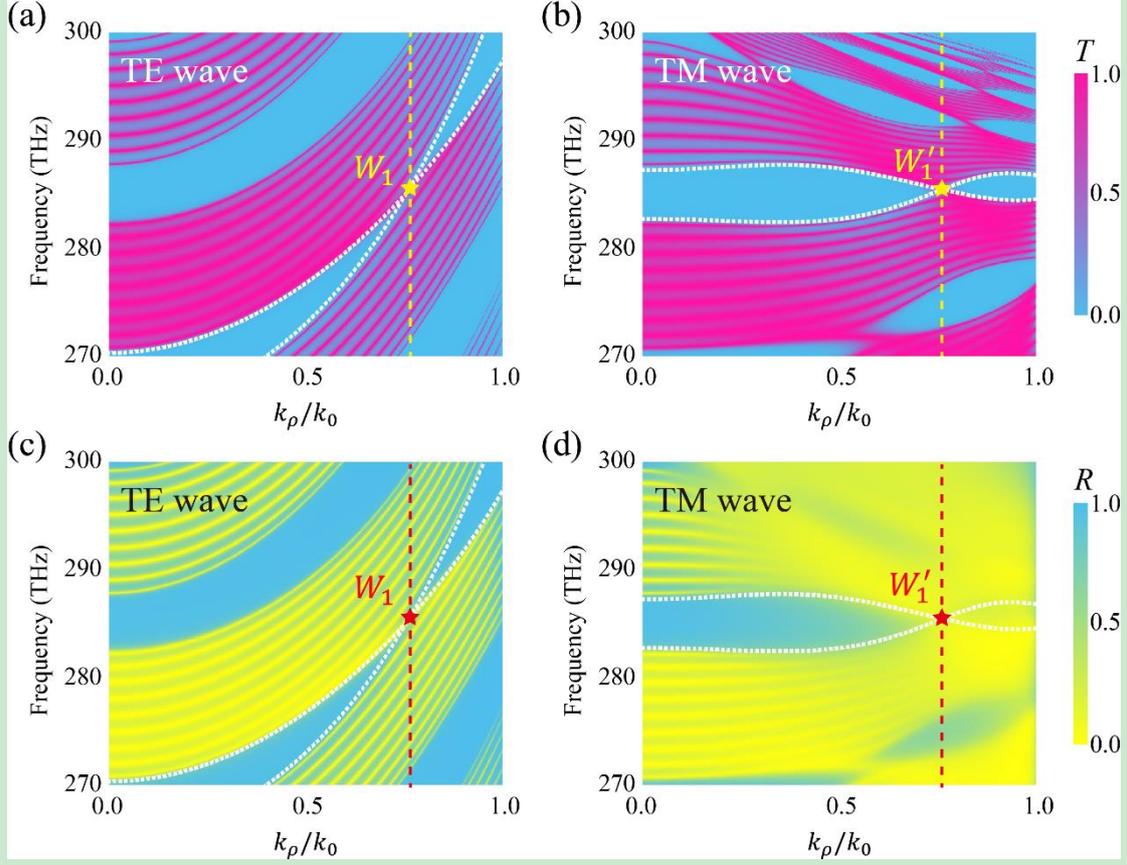

Fig. 7. (a) (b) The transmission spectra of 1D hyper-crystal $[(CD)_{15}B]_{10}$ for TE wave (a) and TM wave (b) when the loss of metal is ignored. At the position of $f_0 = 285$ THz and $k_\rho/k_0 = 0.76$, the type-II WNLS for TE wave and type-I WNLS for TM wave are marked by $W_1$ and $W_1'$, respectively. The band edges for TE and TM waves correspond to $k_z = 0$ and $k_z = \pi/\Lambda$, which are marked by the white dotted lines for see. (c) (d) Similar to (a) (b), but for the reflectance spectra of the 1D hyper-crystal $[(CD)_{15}B]_{10}$ for TE wave (c) and TM wave (d) considering the loss of metal.

For the 1D hyper-crystal which does not satisfy the ideal degenerate condition, that is $m_1 \neq m_2$, $P_1^{TE}/m_1 = P_1^{TM}/m_2$ should be taken into account for realizing the hybrid DNLS. As a result, the parameter groups meeting the degeneration become $(P_1^{TE}d_A/m_1, P_2d_B/n_1) = (P_1^{TM}d_A/m_2, P_2d_B/n_1)$. Here, $m_1 = n_1 = 7$ and $m_2 = 8$. Based on the transfer-matrix method, the calculated transmittance spectra of the 1D hyper-crystal $[(CD)_{15}B]_{10}$ for different normalized wave vector $k_\rho/k_0$ are shown in Figs. 7(a) and 7(b) for TE and TM waves, respectively. It should be noted that in order to show the transmission spectra clearly, we ignore the loss of metal in Figs. 7(a) and 7(b). Although WNLS appears at the same position, $f_0 = 285$ THz and $k_\rho/k_0 = 0.76$ ($\theta = 50^o$) under different polarizations, the types of WNLS of TE and TM polarization waves are different. Especially, the type-II WNLS for TE wave and type-I WNLS for

TM wave are marked by $W_1$ and $W_1'$, respectively. Therefore, these two WNLSs are significantly different from the cases with ideal degenerate conditions shown in Figs. 4 and 5. In fact, in the lossy case considering the loss of metal, the transmission spectra are no longer clear due to the absorption, but similar results are illustrated from the reflectance spectra in Fig. 7(c) and 7(d), where $P_1^{TE} = 7$, $P_1^{TM} = 8$, and $P_2 = 7$. Recently, the non-Hermitian property associated with Dirac point by considering the loss of the system has also attracted people's great attention [86]. Here, the real parts of $\varepsilon_{//}$ and $\varepsilon_\perp$ are about ten times larger than their imaginary parts, so the loss can be nearly ignored.

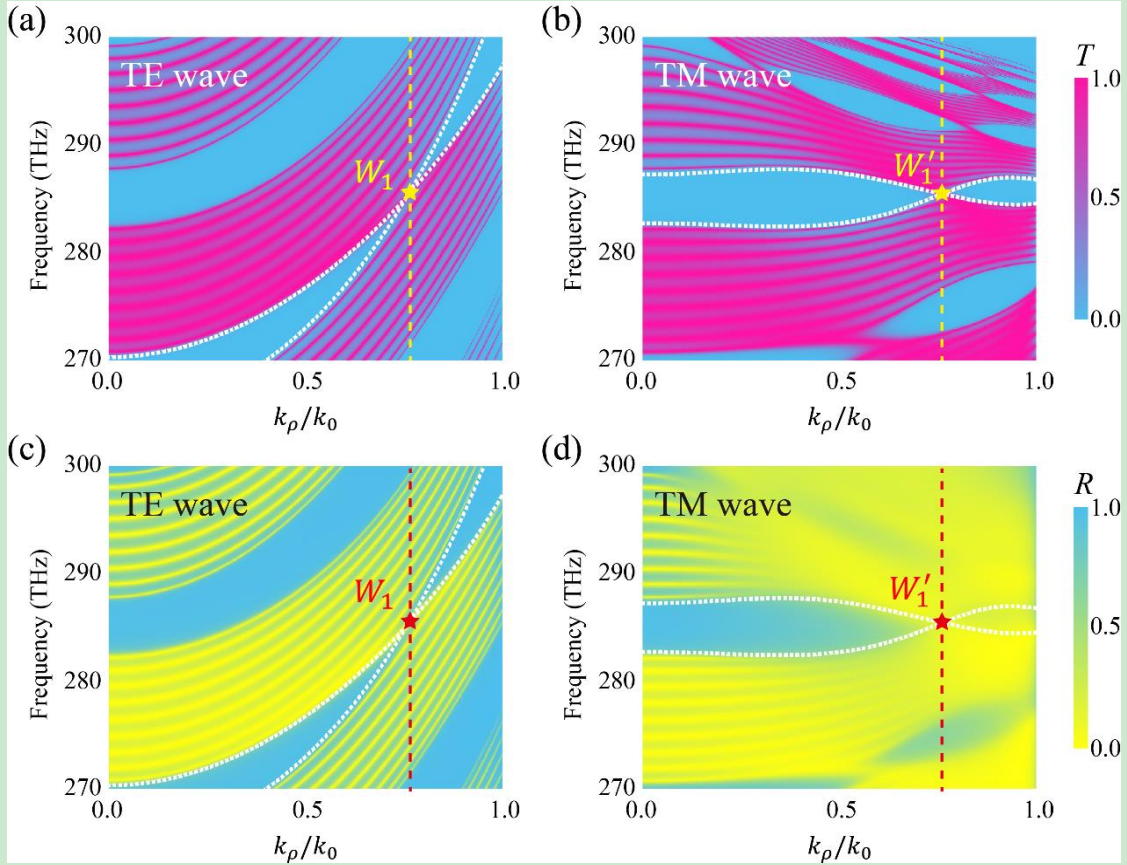

Fig. 8. Similar to Fig. 7, but for the transmittance spectra (lossless structure) and reflectance spectra (lossy structure) of 1D hyper-crystal $[(CD)_{30}B_2]_{10}$. The band edges for TE and TM waves correspond to $k_z = 0$, which are marked by the white dotted lines for see. For TE wave with $k_\rho/k_0 = 0.76$, the WNLS at $W_1$ is marked at the frequency of $f_0 = 285$ THz. At $f_0 = 285$ THz, three WNLS corresponds to $k_\rho/k_0 = 0.46, 0.76$, and $0.96$ are marked by $W_1$, $W_2$, and $W_3$, respectively.

Moreover, after considering the period of HMM layer $A$ and thickness of dielectric layer $B$ are doubled, the hybrid DNLS of the 1D hyper-crystal $[(CD)_{30}B_2]_{10}$ also maintained as $P_1^{TE} = 14$, $P_1^{TM} = 16$, and $P_2 = 14$. In this structure, the transmittance spectra of the lossless case and the reflectance spectra of the lossy case are shown in Fig. 8. It can be seen that the degenerate points for TE and TM waves are maintained, and the type-II WNLS for TE waves and type-I WNLS for TM waves forming the hybrid DNLS [$f_0 = 285$ THz and $k_\rho/k_0 = 0.76$ ($\theta = 50°$)]. Interestingly, at the frequency of $f_0 = 285$ THz, additional WNLSs for the TM waves can be found, which are marked by $W_2$ and $W_3$ at $k_\rho/k_0 = 0.46$ ($\theta = 27°$) and $k_\rho/k_0 = 0.96$ ($\theta = 74°$), respectively. Here, $W_2$ ($W_3$) corresponds to $P_1^{TM} = 15$ (17) and $P_2 = 15$ (13). As a result, multiple concentric Weyl nodal rings can be realized for the TM wave in the 1D hyper-crystal.

At last, we distinguish above introduced two hyper-crystals from the band structures in Fig. 9. The wave vector in the *xoy* plane is $k_\rho = \sqrt{k_x^2 + k_y^2} = 0.76k_0$. The band structure of the 1D hyper-crystal [$(CD)_{15}B]_{10}$ is shown in Fig. 9(a). The results of TE and TM waves are marked by blue and red lines, respectively. We can see that two WNDLs are not coincident on the same $k_z$. The degenerate points for TE and TM correspond to $k_z = 0$ and $k_z = \pi/\Lambda$, respectively. However, two WNDLs are totally overlapped on the same $k_z = 0$ for the 1D hyper-crystal [$(CD)_{30}B_2]_{10}$, as shown in Fig. 9(b). As a result, the novel photonic hybrid DNLS is demonstrated for the structure [$(CD)_{30}B_2]_{10}$ in Fig. 9(b).

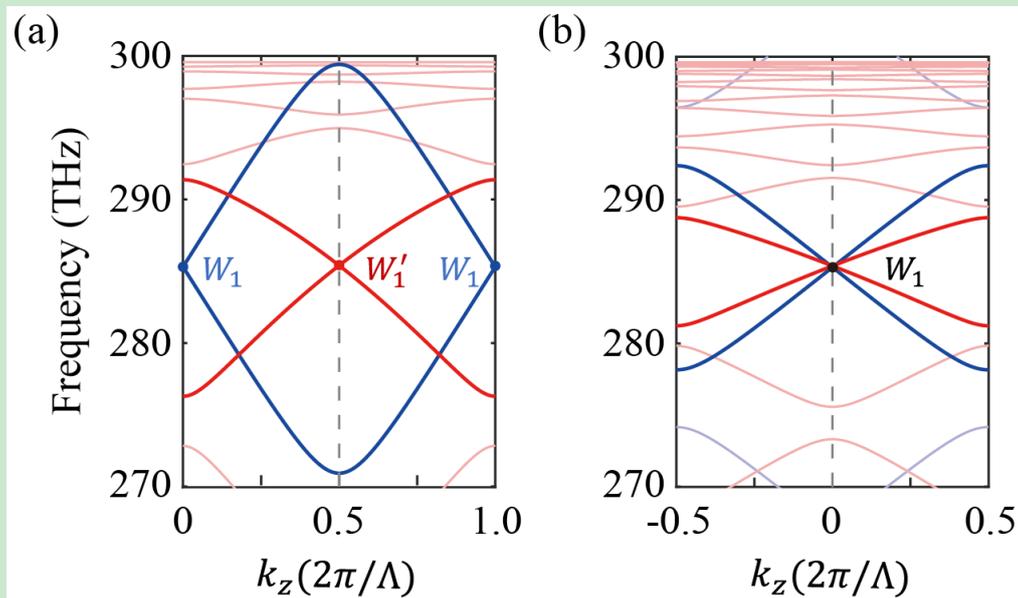

Fig. 9. Band structures of the 1D hyper-crystals $[(CD)_{15}B]_{10}$ (a) and $[(CD)_{30}B_2]_{10}$ (b) for two kinds of polarized waves. The results of TE and TM waves are marked by blue and red lines, respectively. The hybrid DNLSs are formed at $f_0 = 285$ THz. The wave vector in the *xoy* plane is $k_\rho = \sqrt{k_x^2 + k_y^2} = 0.76 k_0$.

## IV. DISCUSSION

In summary, we reveal the various photonic DNLSs and the associated TPT in 1D hyper-crystal, in which the HMM is composed of two types of single-negative media. Moreover, considering the widely used electric HMM composed of metal/dielectric stacks, we observe the novel hybrid DNLS in the 1D hyper-crystal. Our results not only demonstrate various DNLSs in optical regime, but also provide a powerful platform for study on relevant Dirac and Weyl physics. Additionally, the photonic DNLS realized by two types of perpendicularly polarized waves may be used to the polarization independent angle filters.

## ACKNOWLEDGMENT

This work was supported by the National Key R&D Program of China (Grant No. 2021YFA1400602), the National Natural Science Foundation of China (NSFC; Grant Nos. 12004284 and 61621001), the Shanghai Chenguang Plan (No.21CGA22), and the Fundamental Research Funds for the Central Universities (Grant No. 22120210579).

**APPENDIX A: DEGENERATE CONDITIONS OF 1D HYPER-CRYSTAL**

In the main text, the effective electromagnetic parameters of anisotropic HMM layer *A* are $\varepsilon_{Ax} = \varepsilon_{Ay} = \varepsilon_{A\rho} \neq \varepsilon_{Az}$ and $\mu_{Ax} = \mu_{Ay} = \mu_{A\rho} \neq \mu_{Az}$, where subscripts $\rho$ and $z$ indicate components parallel and perpendicular to the anisotropy axis, respectively. Therefore, in order to show clearly, $\varepsilon_{A\rho}$ and $\varepsilon_{Az}$ are expressed by $\varepsilon_\perp$ and $\varepsilon_{//}$, respectively. The wave vector along the direction of radial is indicated as $k_{A\rho} = \sqrt{k_{Ax}^2 + k_{Ay}^2} = k_0 \sin\theta$, where $\theta$ is the incident angle, and $k_0 = \omega/c$ is the wave vector in vacuum. The band structure of the 1D PC is given by

$$\cos(k_z\Lambda) = \cos(k_{Az}d_A)\cos(k_{Bz}d_B) - \frac{1}{2}\left(\frac{\eta_{Az}}{\eta_{Bz}} + \frac{\eta_{Bz}}{\eta_{Az}}\right)\sin(k_{Az}d_A)\sin(k_{Bz}d_B), \quad (A1)$$

where $k_z$ denotes the Bloch wave vector. $\Lambda = d_A + d_B$ is length of the unit cell of 1D PC. $\eta_{iz}$, $k_{iz}$ and $d_i$ are the impedance, the wave vector in $z$ direction, and the thickness of the layer $i$ ($i = A, B$), respectively. Worth mentioning, the polarization of the incident waves should be taken into account. Especially, $\eta_{iz}^{TE} = \frac{\mu_{ix}}{n_{iz}}$, $k_{iz}^{TE} = k_0\tilde{n}_i^{TE}$, and $\tilde{n}_i^{TE} = \sqrt{\varepsilon_{iy}\mu_{ix} - \mu_{ix}k_{i\rho}^2/\mu_{iz}k_0^2}$ ($\eta_{iz}^{TM} = \frac{n_{iz}}{\varepsilon_{ix}}$, $k_{iz}^{TM} = k_0\tilde{n}_i^{TM}$, $\tilde{n}_i^{TM} = \sqrt{\varepsilon_{ix}\mu_{iy} - \varepsilon_{ix}k_{i\rho}^2/\varepsilon_{iz}k_0^2}$) for the TE (TM) wave. In addition, for dielectric layer $B$, $\varepsilon_{B\rho} = \varepsilon_{Bz}$, $\mu_{B\rho} = \mu_{Bz}$ thus $k_{Bz}^{TE} = k_{Bz}^{TM}$. For the 1D hyper-crystal, the band degeneracy will occur at the special point ($f$, $\theta$) in the conditions of $k_{Az}^{TE} = m_1\pi$, $k_{Az}^{TM} = m_2\pi$ and $k_{Bz}^{TE} = k_{Bz}^{TM} = n_1\pi$, where $\{m_1, m_2, n_1\} \in \mathbb{N}^+$ [19]. For the TE wave, $(m_1 + n_1)\ mod\ 2 = 0$ corresponds to the band edges of $k_z = 0$, while $(m_1 + n_1)\ mod\ 2 = 1$ corresponds to $k_z = \pi/\Lambda$. Similarly, for the TM wave, $(m_2 + n_1)\ mod\ 2 = 0$ and $(m_2 + n_1)\ mod\ 2 = 1$ correspond to the band edges of $k_z = 0$ and $k_z = \pi/\Lambda$, respectively. The degenerate conditions of 1D hyper-crystal can be expressed by the ratio of optical path in layers $A$ and $B$

$$\alpha^{TE} = \frac{\tilde{n}_A^{TE}d_A}{\tilde{n}_B^{TE}d_B} = \frac{m_1}{n_1} \in \mathbb{Q}, \text{ and } \alpha^{TM} = \frac{\tilde{n}_A^{TM}d_A}{\tilde{n}_B^{TM}d_B} = \frac{m_2}{n_1} \in \mathbb{Q}. \quad (A2)$$

Especially, for the TE waves, the $(m_1 + n_1)^{th}$ band and the $(m_1 + n_1 + 1)^{th}$ band degenerate at

$$f_{m_1+n_1}^{TE} = \frac{c(m_1+n_1)}{2(\tilde{n}_A^{TE}d_A + \tilde{n}_B^{TE}d_B)}. \quad (A3)$$

Similarly, for the TM waves, the $(m_2 + n_1)^{th}$ band and the $(m_2 + n_1 + 1)^{th}$ band degenerate at

$$f_{m_2+n_1}^{TM} = \frac{c(m_2+n_1)}{2(\tilde{n}_A^{TM}d_A + \tilde{n}_B^{TM}d_B)}, \quad (A4)$$

where $f_{m_1+n_1}^{TE} = f_{m_2+n_1}^{TM} = f_0$.

**APPENDIX B: EM HMM MEETING THE IDEAL DEGENERATION CONDITION $m_1 = m_2$**

For the EM HMM with $\tilde{n}_A^{TE}(f_0) = \tilde{n}_A^{TM}(f_0)$, the relation of effective permittivity and permeability can be expressed as

$$\mu_\perp/\mu_{//} = \varepsilon_\perp/\varepsilon_{//}. \tag{B1}$$

First, we consider that the EM HMM is mimicked by subwavelength stacks as $(CD)_S$. Based on the EMT, the effective electromagnetic parameters are given by [70]

$$\varepsilon_\perp = \varepsilon_{Ax} = \varepsilon_{Ay} = \sum_j f_j \varepsilon_j, \quad \varepsilon_{//} = 1/\sum_j (f_j/\varepsilon_j),$$
$$\mu_\perp = \mu_{Ax} = \mu_{Ay} = \sum_j f_j \mu_j, \quad \mu_{//} = 1/\sum_j (f_j/\mu_j), \tag{B2}$$

where $f_j = d_j/\sum_j d_j$ ($j = C, D$) is the filling factor. Then we can obtain the simultaneous equations

$$\begin{cases} (f_C \varepsilon_C + f_D \varepsilon_D)(\frac{f_C}{\varepsilon_C} + \frac{f_D}{\varepsilon_D}) = (f_C \mu_C + f_D \mu_D)(\frac{f_C}{\mu_C} + \frac{f_D}{\mu_D}) \\ s.t. f_C + f_D = 1 \end{cases}. \tag{B3}$$

According to Eq. (B3), $\frac{\varepsilon_C}{\varepsilon_D} = \frac{\mu_C}{\mu_2}$ or $\frac{\varepsilon_C}{\varepsilon_D} = \frac{\mu_D}{\mu_C}$ should be satisfied, which is not impractical for realizing the photonic DNLSs. This limitation can be solved by considering the three-layered unit cell in the 1D hyper-crystal as $(CDE)_S$. Especially, layers $C$, $D$, and $E$ correspond to ENG media, MNG media and dielectric. In this case, Eq. (B3) becomes

$$\begin{cases} (f_C \varepsilon_C + f_D \varepsilon_D + f_E \varepsilon_E)\xi = (f_C \mu_C + f_D \mu_D + f_E \mu_E)\zeta \\ s.t. f_C + f_D + f_E = 1 \end{cases}, \tag{B4}$$

where $\xi = \frac{f_C}{\varepsilon_C} + \frac{f_D}{\varepsilon_D} + \frac{f_E}{\varepsilon_E}$, and $\zeta = \frac{f_C}{\mu_C} + \frac{f_D}{\mu_D} + \frac{f_E}{\mu_E}$. Compared with Eq. (B3) and Eq. (B4), a new variable $f_E$ is introduced, which provides a new degree of freedom to realize the EM HMM. For the fixed frequency $f_0 = 190$ THz, $\varepsilon_{//}/\varepsilon_\perp - \mu_{//}/\mu_\perp$ as the functions of the filling factor of layer $C$ and layer $D$ are shown by the colormap in Fig. 10. The unit cell of the $(CDE)$ becomes single dielectric layer $E$, single ENG layer $C$, and single MNG layer $D$ at $O$, $Q$, and $P$ point in this triangle phase diagram of the effective electromagnetic parameters. Two axes $OP$ and $OQ$ represent the hyper-crystal with the unit cell of (CE) and (DE), respectively. Especially, $\varepsilon_{//}/\varepsilon_\perp - \mu_{//}/\mu_\perp = 0$ meeting Eq. (B1) is marked by the black dotted line. From Fig. 10, the condition of $\varepsilon_{//}/\varepsilon_\perp - \mu_{//}/\mu_\perp = 0$ can be obtained in the unit cell with three layers with Eq. (B4), instead of the hyper-crystal $(CD)_S$ with Eq. (B3) owning to the black dotted line doesn't fall on two axes $OP$ and $OQ$ other than the origin point. Therefore, the unit cell of the $(CDE)$ is needed for designing the EM HMM.

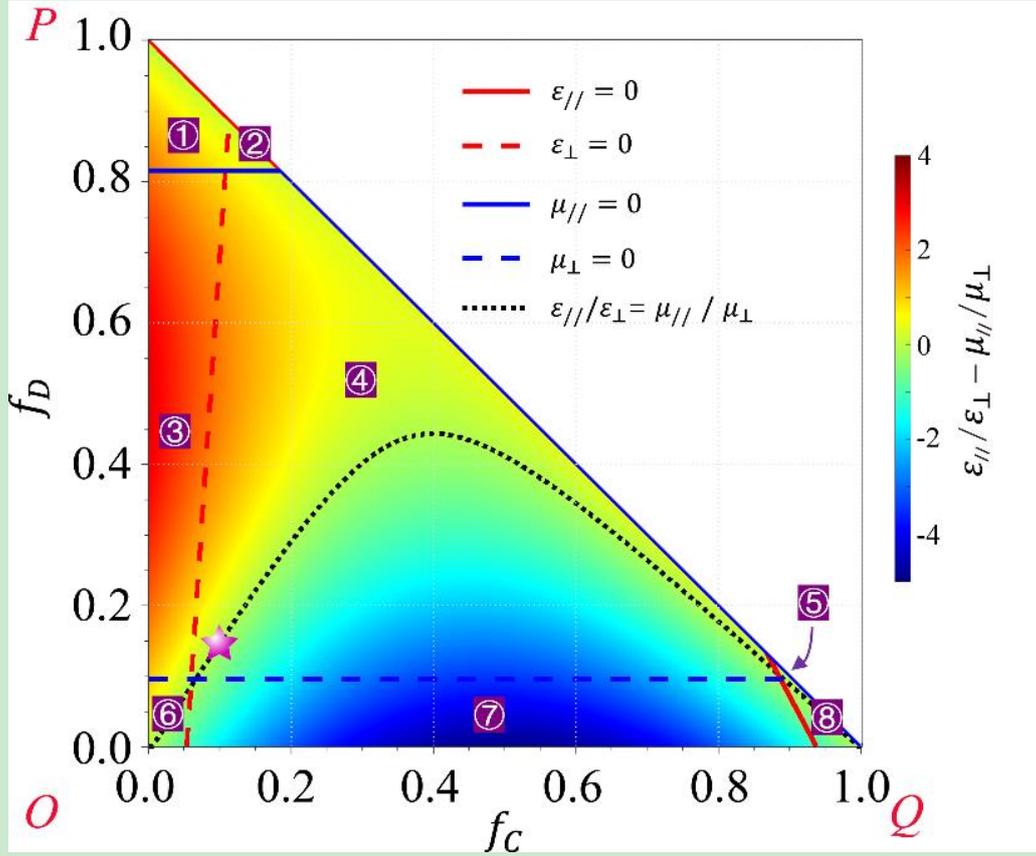

Fig. 10. The phase diagram of the effective electromagnetic parameters as the functions of $f_C$ and $f_D$. The colormap represents the value of $\varepsilon_{//}/\varepsilon_\perp - \mu_{//}/\mu_\perp$. $\varepsilon_{//}/\varepsilon_\perp - \mu_{//}/\mu_\perp = 0$ is shown by the black dotted line. $\varepsilon_{//} = 0$ ($\mu_{//} = 0$) and $\varepsilon_\perp = 0$ ($\mu_\perp = 0$) are represented by the red (blue) solid line and dashed line, respectively. Eight subspaces are divided by the zero points of the anisotropic electromagnetic parameters.

Moreover, $\varepsilon_{//} = 0$ ($\mu_{//} = 0$) and $\varepsilon_\perp = 0$ ($\mu_\perp = 0$) are represented by the red (blue) solid line and dashed line, respectively. These lines divide the triangle phase diagram into eight subspaces. The details of electromagnetic parameters in different subspaces are summarized in Table 2. Especially, 'None' denotes the effective media without real wave vectors, 'Type I' and 'Type II' means two types of HMMs, and the 'Elliptic' corresponds to the anisotropic dielectric with elliptic dispersion.

| subspace | Effective parameters | | | | Dispersion | |
|---|---|---|---|---|---|---|
| | $\varepsilon_\perp$ | $\varepsilon_{//}$ | $\mu_\perp$ | $\mu_{//}$ | TM | TE |
| ① | + | + | - | - | None | None |
| ② | + | - | - | - | Type II | None |
| ③ | + | + | + | - | Elliptic | Type I |
| ④ | + | - | + | - | Type I | Type I |
| ⑤ | - | - | + | - | None | Type II |
| ⑥ | + | + | + | + | Elliptic | Elliptic |

| ⑦ | + | - | + | + | Type I | Elliptic |
| ⑧ | - | - | + | + | None | None |

**Table 2.** The ranges of effective parameters and dispersions of the subspaces shown in Fig. 10.

More attention is paid to the phases supporting both TE and TM polarized waves. Figure 11 indicates iso-frequency contour in the subspace ③, ④, ⑥, and ⑦, where the blue (red) solid lines represent the TE (TM) waves. Importantly, the blue (red) dashed line implies the process of magnetic (electrical) Lifshitz transition [70]. The EM HMM studied in this work belongs to the subspace ④. In this work, the filling factor of layer $C$, $D$, and $E$ layers is selected as $f_C = 0.1$, $f_D = 0.15$, and $f_E = 0.75$, respectively., which is marked by the star near the black dotted line for see. Figure 3(c) in the main text proves the effectiveness of condition given in Eq. (B1). The effective refractive indexes of the TE (the blue solid line) and TM (the red triangles) waves in Fig. 3(c) are almost completely coincident independent on the incident angle at the degenerate frequency $f_0 = 190.4$ THz.

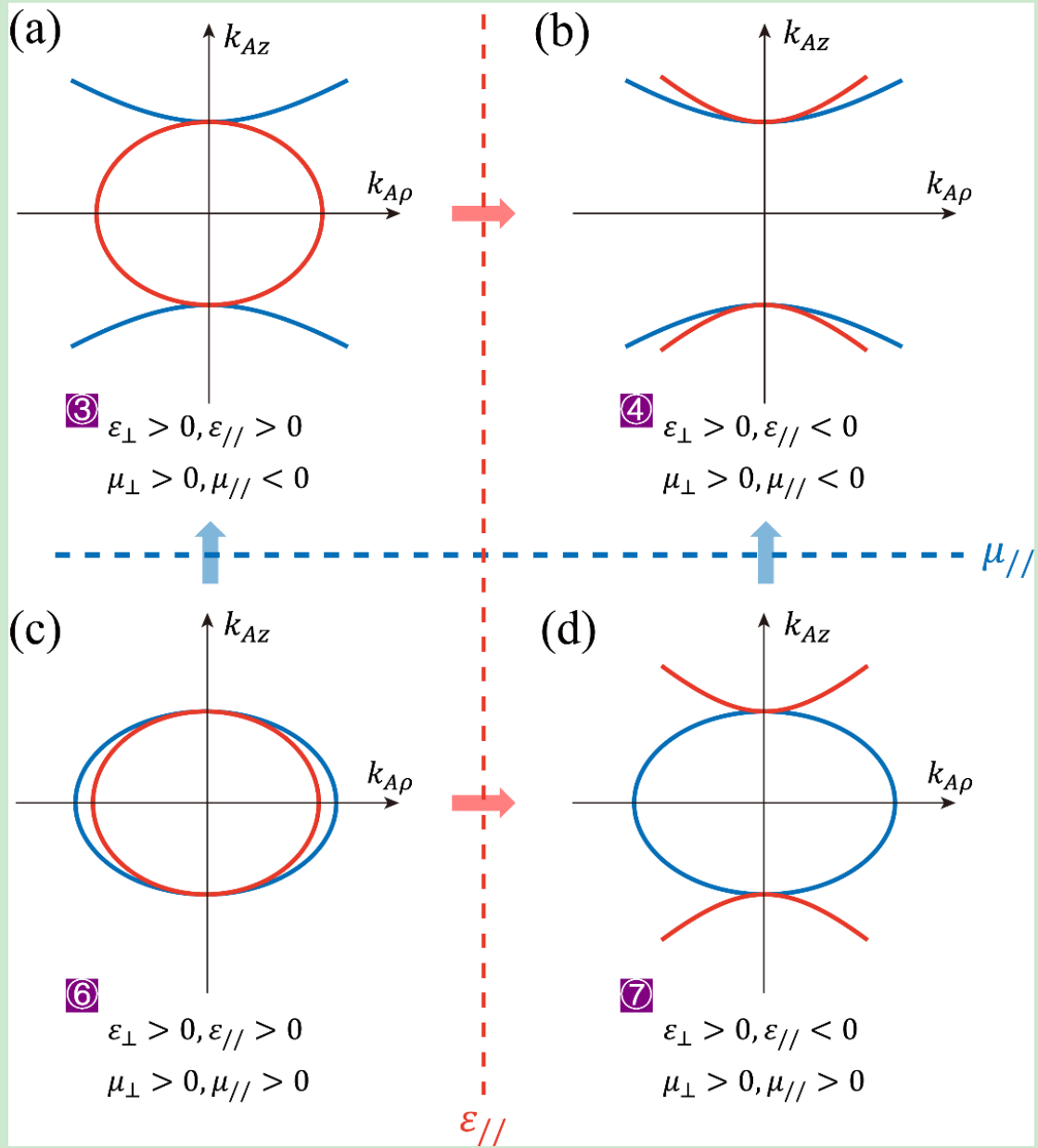

Fig. 11. Iso-frequency contours of the layer $A$ in the subspaces ③, ④, ⑥, and ⑦, where the blue (red) solid lines represent the TE (TM) wave. The blue (red) dashed line implies the process of magnetic (electrical) Lifshitz transition.

**APPENDIX C: THE EFFECTIVENESS OF EMT FOR HMM**

In this section, the effectiveness of the EMT for EM HMM and electric HMM in the main text is demonstrated. First, for the EM HMM with effective parameters under ideal degeneration condition $m_1 = m_2$, the transmission spectra of 1D hyper-crystal $[AB]_{100}$ for the TE and TM waves are shown in Fig. 12(a) and 12(b), respectively. Here the thickness of layer $A$ and $B$ are $d_A = 624\ nm\ (S = 4)$ and $d_B = 352\ nm$, respectively. The other parameters are the same as that in Fig. 3. It can be clearly seen that the adjacent bands cross each other at $f_0 = 190$ THz and $k_\rho/k_0 = 0.66$ ($\theta = 40^o$), which is shown by the yellow star. In addition, the band edges with $k_z = \pi/\Lambda$ are marked by white dotted lines. This degenerate point corresponds to $m_1 = m_2 = 2$ and $n_1 = 1$. From Figs. 12(a) and 12(b), we can see the 1D hyper-crystal belongs to a type-I DNLS, which is same as the results shown in Figs. 3(d) and 3(e). For comparison, based on the transfer-matrix method, figures 12(c) and 12(d) directly give the transmission spectra of 1D hyper-crystal 1D hyper-crystal [(CDE)$_4$B]$_{100}$ for the TE and TM waves, which meet well with the results of EMT in Figs. 12(a) and 12(b). Therefore, the EMT for the EM HMM in the 1D hyper-crystal is reasonable. On the other hand, in Fig. 12, $P_1 = 2$ and $P_2 = 1$ which are half of the corresponding parameters in Fig. 3. As a result, the band edges represented by $k_z = 0$ in Figs. 3(d) and 3(e) are changed to $k_z = \pi/\Lambda$ in Fig. 12.

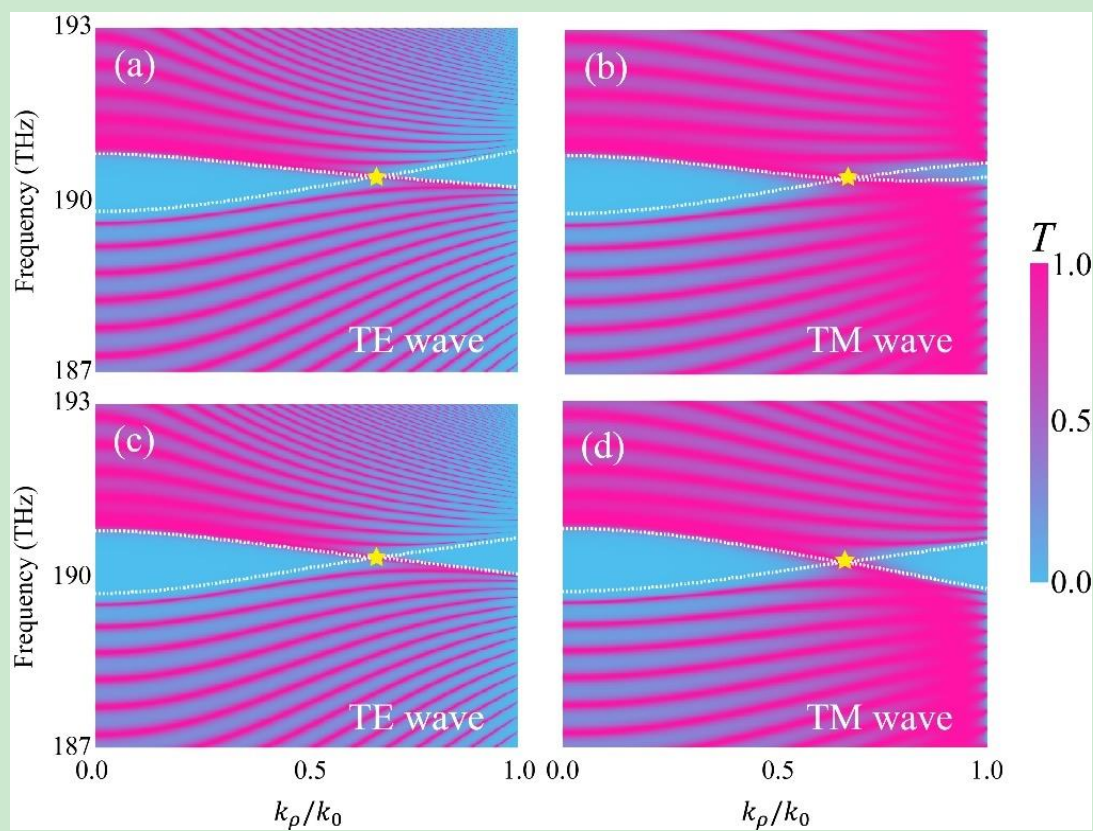

Fig. 12. (a) (b) Under EMT, the transmission spectra of 1D hyper-crystal [$AB$]$_{100}$ with EM HMM for the TE wave (a) and TM wave (b). (c) (d) The transmission spectra of 1D hyper-crystal [(CDE)$_4$B]$_{100}$ with EM HMM for the TE wave (c) and TM wave (d). The band edges with $k_z = \pi/\Lambda$ are marked by white dotted lines.

Considering the 1D hyper-crystal with electric HMM under the condition $m_1 \neq m_2$, we also study the effectiveness of the EMT for the photonic DNLS. Based on the EMT, the transmission and reflectance spectra of 1D hyper-crystal $[AB]_{10}$ for TE and TM waves are shown in Fig. 13. Compared the lossless case in Fig. 13(a) and lossy case in Fig. 13(b) for TE wave, we can clearly see that the loss of the system will blur the transmission spectra. However, the photonic WNLS also can be clearly observed in the reflectance spectra, as shown in Fig. 13(c). Similar results are found for TM wave in Figs. 13(d)-(f). Especially, the calculated results under EMT in Fig. 13 are meet well with the results directly based on the transfer-matrix method for $[(CD)_{15}B]_{10}$ in Fig. 7. Therefore, the effectiveness of the EMT for EM HMM and electric HMM in the main text is demonstrated in Figs. 12 and 13.

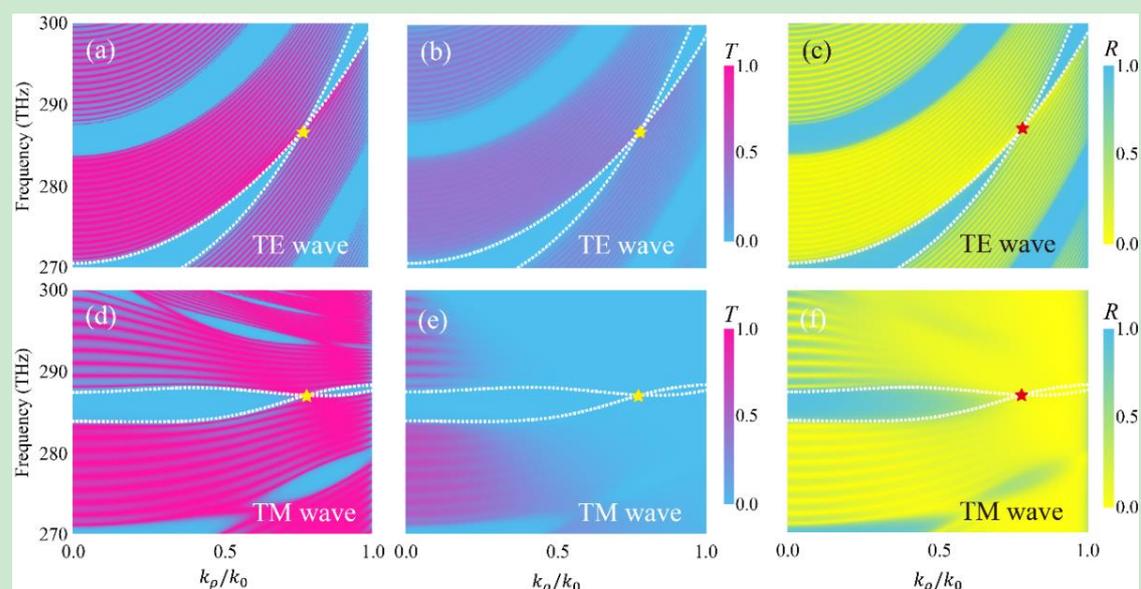

Fig. 13. (a) (b) The transmission spectra of 1D hyper-crystal $[AB]_{10}$ with electric HMM for the lossless and lossy cases under TE wave. (c) The reflectance spectra of 1D hyper-crystal $[AB]_{10}$ for the lossy case. (d)-(f) Similar to (a)-(c), but for the TM wave. The band edges for TE and TM waves correspond to $k_z = 0$ and $k_z = \pi/\Lambda$, which are marked by the white dotted lines for see.

At last, the refractive indexes ($n$) of the layer $B$: $KTiOPO_4$ and the layer $C$: $ZnSiAs_2$ in the section III are shown respectively by the cyan and orange lines in Fig. 14. It can be seen that the refractive indexes of layer $B$ and layer $C$ are close to 1.47 and 3.3, and there is slight dispersion near the working frequency $f_0 = 285$ THz.

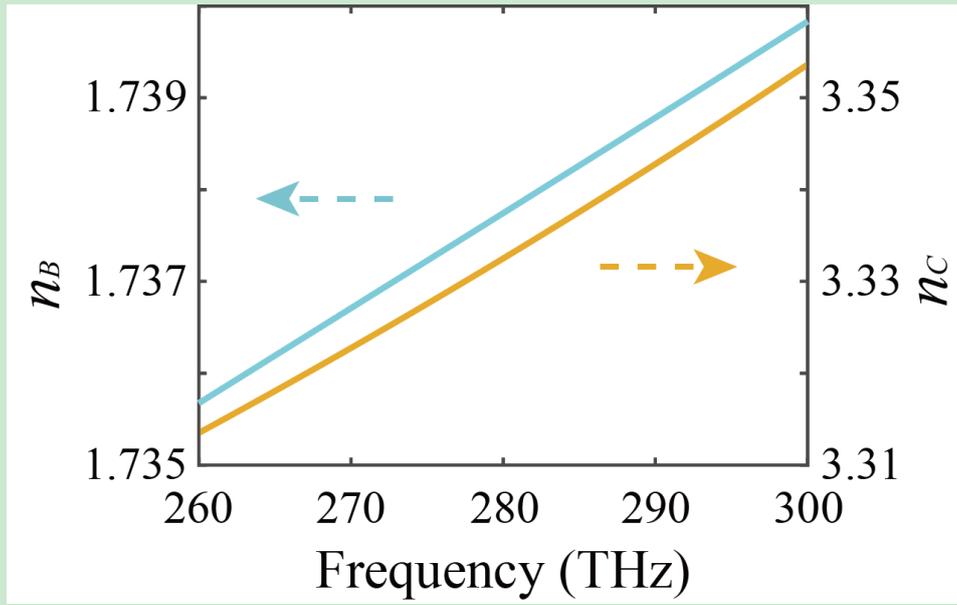

Fig. 14. The refractive indexes ($n$) of the layer $B$: $KTiOPO_4$ and the layer $C$: $ZnSiAs_2$.